

\parindent=15pt
\parskip=0pt plus 1pt minus 0.5pt
\baselineskip=12pt plus 0.5pt minus 0.5 pt

\def\pd{\partial}

\centerline{}

\medskip
\centerline{\bf EFFECT OF ADIABATIC DECELERATION ON THE FOCUSED TRANSPORT}
\medskip
\centerline{\bf OF SOLAR COSMIC RAYS}

\bigskip
\centerline{D.~RUFFOLO}
\medskip
\centerline{Department of Physics, Chulalongkorn University, Bangkok 10330,
Thailand}
\medskip
\centerline{\it Accepted for publication in the Astrophysical Journal}

\bigskip
\bigbreak
\medskip
\centerline{ABSTRACT}
\medskip
In the framework of focused transport theory, adiabatic deceleration arises
from adiabatic focusing in the solar wind frame and from differential solar
wind convection.  An explicit formula is given for the deceleration of
individual particles as a function of the pitch angle.  Deceleration and
other first-order effects of the solar wind, including convection, are
incorporated into a numerical code for simulating the transport of energetic
particles along the interplanetary magnetic field.  We use this code to
model the transport of solar flare protons.  We find that including
deceleration can increase the decay rate of the near-Earth intensity by 75\%
more than would be expected based on advection from higher momenta, due to
an interplay with diffusive processes.  Improved response functions are
derived for the impulsive injection of particles near the Sun, and it is
found that neglecting deceleration leads to incorrect estimates of the
scattering mean free path based on the intensity decay alone, especially for
lower-energy particles.

\bigbreak
\medskip
\centerline{1. \  INTRODUCTION}
\medskip
The study of solar cosmic rays is marked by an incomplete
understanding of the mechanisms for their acceleration and transport
through the solar corona.  One aspect of their behavior which is
relatively well understood is their transport through the
interplanetary medium.  Because of this, one can attempt to perform
numerical simulations of interplanetary transport, by which to
deconvolve the measurements of solar cosmic rays by interplanetary
spacecraft and derive their injection profile near the sun.

Recently, several authors have presented detailed numerical simulations of
the focused transport of solar flare particles in the interplanetary medium
(e.g., Earl 1976; Bieber 1977; Ng \& Wong 1979;
K\'ota et al.\ 1982; Earl 1987; Ruffolo 1991; Pauls \& Burger 1993).
While these authors used a variety of numerical
methods, they all treated particle transport in a framework
in which particles follow helical orbits about the interplanetary magnetic
field, subject to pitch-angle scattering and magnetic ``focusing'' or
``mirroring''; we will refer to this as the
framework of focused transport.  These numerical
studies have thus treated the particle distribution function, $F$, as
depending on the time since the flare occurrence, $t$, the pitch-angle
cosine, $\mu$, and the distance along the interplanetary field, $z$.

One transport process which was not included in these previous
treatments is that of adiabatic deceleration, although its effect on
solar cosmic-ray distributions has been clearly demonstrated
(e.g., Murray, Stone, \& Vogt 1971).  This process is
commonly described as the adiabatic cooling of a isotropic cosmic ray ``gas''
in an expanding flow.  It was proposed by Singer (1958) and Singer, Laster,
\& Lenchek (1962) that the deceleration of
cosmic rays in expanding clouds of ``magnetized solar plasma''
could account for temporary Forbush decreases in cosmic ray
intensity.  Parker (1965) considered the expanding solar wind, and
included the effects of deceleration, spatial diffusion, and convection
in a Fokker-Planck equation for the cosmic ray distribution as a
function of the radius and kinetic energy.  Modified versions of
this are still commonly used, e.g., in the study of solar modulation
of galactic cosmic rays.   Note that these treatments assume
a nearly isotropic directional distribution.

On the other hand, Jokipii (1966) used a Fokker-Planck equation
incorporating pitch-angle scattering and streaming to explain the
cosmic ray distribution as a function of the distance along the
magnetic field and the cosine of the pitch angle, which is defined as
the angle between the particle's velocity and the outward tangent to
the average magnetic field.  Once the effect of focusing was added
(Earl 1976), this grew into the above mentioned framework of focused
transport, which can explicitly model the anisotropy that is often
observed in the cosmic ray distribution immediately after a solar
flare event.

Thus for the most part, descriptions of solar cosmic-ray transport
have included either momentum-changing solar wind effects (such as
deceleration and convection) or pitch-angle dependent effects, but not both.
Notable exceptions include the theoretical treatments of Skilling
(1971, 1975), Luhmann (1976), Webb \& Gleeson (1979), Earl (1984),
and Schlickeiser (1989).  To the author's knowledge, no numerical
simulations including both types of effects have been reported before.

In this work, we present an analytic treatment and numerical
simulations of the interplanetary transport of energetic particles,
including both an\-iso\-trop\-ic pitch-angle distributions and solar wind
effects such as deceleration and convection.  The latter are incorporated
in the framework of focused transport, which already includes the former.
After a description of adiabatic deceleration in this framework,
and the derivation of explicit formulae for deceleration and other
solar wind effects as a function of pitch angle, we present the equation of
transport in a form which is suitable for numerical solution.
We then describe a newly developed numerical code for the solution of this
equation.  As an application of the code, we examine the results of
simulations of the interplanetary transport of solar cosmic rays.
We find that the late-time decay due to ``switching on'' deceleration can be
75\% faster than one would expect on the basis of momentum advection alone,
because the advection rate is not spatially uniform, which enhances
the decay rate due to spatial diffusion.
Finally, we discuss possible future applications of the numerical
method to the study of solar and galactic cosmic ray transport.

\bigbreak
\centerline{2. \  EQUATION OF TRANSPORT}
\medskip
\centerline{2.1. \ {\it Adiabatic Deceleration}}
\medskip
Many studies of cosmic ray transport, particularly those concerning
solar modulation, have used the formula
$$\langle\dot p'\rangle = -{p'\over 3}\nabla\cdot{\bf v}_{\rm
sw},\eqno(1)$$
where ${\bf v}_{\rm sw}$ is the solar wind velocity and $p'$ is
the particle momentum in the solar wind frame (Parker 1965; Dorman 1965).
However, this average rate of deceleration is only valid when the particles'
pitch-angle distribution is nearly isotropic.  In this section we relax this
assumption, and consider the deceleration of individual particles as a
function of the pitch angle.  Our approach represents an extension of the
focused transport model to include effects of the solar wind to first
order in $(v_{\rm sw}/c)$ and examine those effects by means of numerical
simulations; these effects have also been examined analytically in a more
general context or to higher orders, e.g., by Luhmann (1976), Webb \& Gleeson
(1979), Earl (1984), and Schlickeiser (1989).

In the focused transport model (Earl 1976),
energetic particles are considered to undergo pitch-angle
scattering due to small-scale irregularities in the interplanetary
magnetic field, and ``focusing'' or ``mirroring'' due to the large-scale
divergence or weakening of the field at increasing distances
from the sun.  The latter process systematically decreases the pitch
angle and increases $\mu$, so that $v_\parallel=\mu v$ systematically
increases, where $v_\parallel$ is the component of the particle
velocity moving outward from the sun along the local, average interplanetary
magnetic field line.  In this model the solar wind velocity is generally
neglected.  Both the large-scale and small-scale features of the
field are treated as stationary, and the magnitude of the particle
velocity is conserved by both focusing and scattering.

If, on the other hand, we do not neglect the solar wind velocity, we
are led to consider two reference frames: the fixed frame and the local solar
wind frame (comoving with the solar wind velocity at a given point).
We denote particle velocities in these frames as $\bf v$ and $\bf v'$,
respectively, but for convenience we will always use the position in
the fixed frame, following Jokipii \& Parker (1970).
Because the large-scale structure of the magnetic field is taken
to be stationary in the fixed frame, the process of focusing conserves
$v=|\bf v|$ (see Fig.~1a). Similarly,
small-scale irregularities in the field can be considered to be
frozen in the solar wind to a good approximation (Parker 1965).
Then in the solar wind frame, the process of scattering conserves
the magnitude of the velocity, $v'=|\bf v'|$.

\bigbreak
\medskip
\centerline{2.1.1. \ {\it Radial Magnetic Field}}
\medskip
Now let us examine the effects of scattering and
focusing as viewed in the solar wind frame (Fig.~1b).
For simplicity, we first consider a radially-directed magnetic field
with superimposed small-scale irregularities, assume a constant solar
wind speed, and perform the velocity transformations non-relativistically.
Scattering preserves the magnitude of the velocity, $v'$, or the
distance from the origin in the $v'_\parallel-v'_\perp$ plane.
Thus scattering moves the velocity back and forth along
semicircles centered at the origin (solid lines in Fig.~1b).
Now the fixed frame moves at a velocity $-{\bf v}_{\rm
sw}$ with respect to this frame, as indicated by the point F in
Fig.~1b; therefore, ${\bf v}={\bf v'}+{\bf v}_{\rm sw}$.
Focusing preserves $v$, which is the distance from
point F, and increases $v_\parallel=v'_\parallel+v_{\rm sw}$.
Thus this process moves the velocity to the right along semicircles
centered at F (dotted lines in Fig.~1b); note that this is
equivalent to the $dB/dt$ effect in plasma physics.  However, $v'$ is
not conserved, and in fact {\it systematically decreases} as focusing
inevitably carries the particle velocity closer
to the origin.  As an example, Fig.~2 shows the schematic
trajectory of a particle which alternately undergoes scattering
and focusing, the latter of which always decreases $v'$.

The rate of deceleration for this example can be calculated in
a straightforward way.  Note that
$$
\dot v' = \left.dv'\over d\mu\right|_v \dot\mu. \eqno(2)
$$
The rate of change of $\mu$ in the fixed frame due to focusing,
$\dot\mu$, is given by
$$
\dot\mu = {v\over 2L(z)}(1-\mu^2) \eqno(3)
$$
(Roelof 1969), where
$$
{1\over L(z)}\equiv -{1\over B}{dB\over dz}
$$
and $z$ is the arclength from the sun along the magnetic field.
Substituting $\dot\mu$ into equation~(2), evaluating the derivative, and
simplifying, we find that
$$
\dot v' = -{v_{\rm sw}v'\over 2L(z)}(1-\mu^{\prime 2}).
$$
A relativistic derivation yields a similar formula for the rate of
change of the momentum (neglecting terms of order $(v_{\rm sw}/c)^2$;
normally $v_{\rm sw}/c=0.001$ to 0.002 near the Earth):
$$
\dot p' = -{v_{\rm sw}p'\over 2L(z)}(1-\mu^{\prime 2}). \eqno(4)
$$

For the radial field we are considering,
$$
{v_{\rm sw}\over L(z)} = {2v_{\rm sw}\over r} = \nabla\cdot{\bf v}_{\rm sw},
$$
and therefore
$$
\dot p' = -{p'\over 2}(1-\mu^{\prime 2})\nabla\cdot{\bf v}_{\rm sw}.
$$
Then if the pitch-angle distribution in the solar wind frame is
isotropic, the directional average of $1-\mu^{\prime 2}$ is $2/3$, and
$$
\langle\dot p'\rangle = -{p'\over 3}\nabla\cdot{\bf v}_{\rm sw}.
$$
Therefore, the pitch-angle dependent formula, equation~(4), yields
equation~(1) for the case of an isotropic pitch-angle distribution.

In the derivation presented above, adiabatic deceleration is
a monotonic decrease in the momentum resulting from the transformation of
adiabatic focusing from the fixed frame to the solar wind frame.
However, adiabatic deceleration need not take place in a magnetic
field; it would occur, for example, in a system of particles
colliding with heavy ``scattering centers'' flowing away from the
origin according to the velocity field, ${\bf v}_{\rm sw}$
(Gleeson \& Axford 1967; Jones 1990).  For this case, the motion of a
particle between collisions is along a straight line.
However, since the orientation of ${\bf v}_{\rm sw}$ varies with position,
the parallel component of the particle's velocity changes with time.
Using spherical coordinates, we have
$$
\dot v_r = {v_\theta^2 + v_\varphi^2\over r},
$$
or in terms of components parallel and perpendicular to ${\bf v}_{\rm sw}$,
$$
\dot v_\parallel = {v_\perp^2\over r}.\eqno(5)
$$
This kinematic effect is equivalent to adiabatic focusing
(eq.~[3]), because a magnetic field parallel to
${\bf v}_{\rm sw}$ does not affect $v_\parallel$.
Thus adiabatic focusing (mirroring), or the increase in
$v_\parallel$, can be viewed as a kinematical effect; the magnetic
field merely serves to hinder the perpendicular motion.
Equation (5), when transformed into the solar wind frame, then
leads to adiabatic deceleration, whether or not a magnetic field is present.

\bigbreak
\medskip
\centerline{2.1.2. \ {\it Spiral Magnetic Field}}
\medskip
So far, we have derived a pitch-angle dependent formula for the
deceleration of a particle in a radial magnetic field.
We now derive a similar formula for the case of an Archimedean spiral
field.  For convenience, we can use a fixed frame that
is corotating with the sun, so that solar wind velocity, ${\bf v}^c_{\rm sw}$,
is parallel to the magnetic field at each point (Parker 1958). In particular,
$$\eqalign{
{\bf v}^c_{\rm sw} &= v_{\rm sw}\hat r - \Omega r\sin\theta\hat\varphi\cr
 &= v^c_{\rm sw}\hat z,\cr
}$$
and
$$
v^c_{\rm sw} = v_{\rm sw}\sec[\psi(z)],
$$
where $\Omega$ is the angular velocity of the solar rotation, $\hat
z$ is along the outward tangent to the average magnetic field, and
$\psi(z)$ is the angle between $\hat r$ and $\hat z$ (see Fig.~3).
Thus the corotating fixed frame and the local solar wind frame
have a relative velocity ${\bf v}^c_{\rm sw}$, and since
this is parallel to the magnetic field, we can use the framework of
the previous section (see Fig.~1).

As before, there is a systematic decrease in $p'$, the magnitude of
the particle momentum in the solar wind frame, due to adiabatic
focusing, as given by equation~(4), replacing $v_{\rm sw}$ with $v^c_{\rm
sw}$.  There is also an additional
decrease due to changes in the solar wind velocity, $v^c_{\rm sw}$,
along the magnetic field.  [These two effects are related to the
perpendicular and parallel divergence of ${\bf v}^c_{\rm sw}$,
respectively, and were termed ``betatron deceleration'' and ``an inverse
Fermi effect'' by Webb \& Gleeson (1979).]
Because of changes in ${\bf v}^c_{\rm sw}$,
the local solar wind frames are different at different locations;
we call this effect ``differential convection.''
Consider a particle streaming from point A to point B
on the same field line.  We need to relate $p'_{\rm A}$, the momentum in
the local solar wind frame at point A, to $p'_{\rm B}$ at point B.
Assuming a constant momentum in the fixed frame, and neglecting
terms of order $(v^c_{\rm sw}/c)^2$, we have
$$\eqalign{
p'_{\parallel,{\rm A}} &= p_\parallel - {E\over c^2}v^c_{\rm sw,A}\cr
p'_{\parallel,{\rm B}} &= p_\parallel - {E\over c^2}v^c_{\rm sw,B}\cr
\Delta p'_\parallel &= -{E\over c^2}v_{\rm sw}(\Delta\sec\psi).\cr
}$$
If $z$ is the arclength along the Archimedean field line,  $\Delta
z$ is the distance from A to B, and $\Delta t$ is the travel time, then
$$\eqalign{
\Delta p'_\parallel &= -{E\over c^2}v_{\rm sw}
\left({d\over dz}\sec\psi\right)\Delta z\cr
&= -{E\over c^2}v_{\rm sw}\left(\cos\psi{d\over
dr}\sec\psi\right)v_\parallel\Delta t\cr
\dot p'_\parallel &= -p_\parallel v_{\rm sw}\left(\cos\psi{d\over dr}
\sec\psi\right).\cr
}$$
To examine the effect on $p'$ due to differential convection alone,
we can fix the value of $p'_{\perp}$.  Since $p_\parallel v_{\rm sw} =
p'_\parallel v_{\rm sw}$ to first order in $(v_{\rm sw}^c/c)$, we find that
the rate of deceleration due to changes in $v^c_{\rm sw}$ is
$$\eqalignno{
\dot p' &= {p'_\parallel\over p'}\dot p'_\parallel&\cr
&= -p'v_{\rm sw}\left(\cos\psi{d\over dr}\sec\psi\right)\mu^{\prime2}.
&(6)\cr}
$$
Finally, adding this rate of deceleration to that derived from equation~(4)
yields a total deceleration rate of
$$
\dot p' = -p'v_{\rm sw}\left[{\sec\psi\over 2L(z)}(1-\mu^{\prime 2})
+\cos\psi{d\over dr}\sec\psi\mu^{\prime2}\right], \eqno(7)
$$
which can again be shown to be equivalent to equation~(1) for
an isotropic pitch-angle distribution.  Note that this expression for
$\dot p'$ is equivalent to that used by Skilling (1971).

\bigbreak
\medskip
\centerline{2.2. \ {\it Other Effects of the Solar Wind}}
\medskip
Besides adiabatic deceleration, the solar wind velocity has other
effects on the transport of solar flare particles.  The effects
described in this section are included in the transport code for the
sake of completeness, although some of them represent rather small corrections
for energetic particles.  From now on we will continue to
assume a constant, radial solar wind and an Archimedean
spiral magnetic field.

One well-known effect is solar wind convection (e.g., Parker 1965).
Since the distribution of particles tends to become isotropic in the
solar wind frame due to pitch-angle scattering, solar wind convection
can cause an anisotropy in the distribution observed in the fixed
frame (the Compton-Getting effect; Compton \& Getting 1935; Forman 1970).
Comparing the parallel velocities in the fixed and corotating frames, we have
$$
p_\parallel = \mu'p' + {E\over c^2}v_{\rm sw}\sec\psi,
$$
and the rate of streaming is given by
$$\eqalignno{
\dot z &= v_\parallel = {E'\over E}\mu'v' + v_{\rm sw}\sec\psi&\cr
  &= \mu'v' + \left(1-\mu^{\prime 2}{v^{\prime 2}\over c^2}\right)
     v_{\rm sw}\sec\psi,&(8)\cr
}$$
again neglecting terms of order $(v^c_{\rm sw}/c)^2$ or higher.

In the standard focused transport model, the pitch-angle cosine in the
fixed frame, $\mu$, changes due to adiabatic focusing, as given by
equation~(3).  In the local solar wind frame, this becomes
$$\eqalignno{
\dot \mu' &= \left.d\mu'\over d\mu\right|_p \dot\mu &\cr
&= {v'\over 2L(z)}\left[1+\mu'{v_{\rm sw}\over v'}\sec\psi
-\mu'{v_{\rm sw}v'\over c^2}\sec\psi\right](1-\mu^{\prime2}),
&(9)\cr
}$$
neglecting terms of order $(v^c_{\rm sw}/c)^2$.  Note the
first-order correction factor in brackets.

In addition, $\dot\mu'$ is affected by differential convection,
i.e., changes in the solar wind velocity, $v^c_{\rm sw}$.  Equation~(6)
shows the contribution to $\dot p'$ due to this effect.
Since $p'_\perp$ remains unaffected, we have
\vskip10pt
$$\eqalign{
d(p^{\prime2}_\perp) = 0 &= d[p^{\prime2}(1-\mu^{\prime2})]\cr
&= 2(1-\mu^{\prime2})p'dp' - 2\mu'p^{\prime2}d\mu'\cr
\dot\mu' &= {1-\mu^{\prime2}\over\mu'}{\dot p'\over p'}.\cr
}$$
Using equation~(6), we find that to first order in $(v^c_{\rm sw}/c)$,
the rate of change of $\mu'$ due to changes in $v^c_{\rm sw}$ is
$$
\dot\mu' = -v_{\rm sw}\left(\cos\psi{d\over dr}\sec\psi\right)
\mu'(1-\mu^{\prime2}).
$$
Combining this with the rate due to focusing, equation~(9), we find
that the total rate of change is
$$
\dot\mu' = {v'\over 2L(z)}
\left[1+\mu'{v_{\rm sw}\over v'}\sec\psi
-\mu'{v_{\rm sw}v'\over c^2}\sec\psi\right]
(1-\mu^{\prime2})
$$
$$
-v_{\rm sw}\left(\cos\psi{d\over dr}\sec\psi\right)
\mu'(1-\mu^{\prime2}).\eqno(10)
$$

\bigbreak
\medskip
\centerline{2.3. \ {\it Modified Equation of Focused Transport}}
\medskip
In the previous sections, we considered the transport
of particles in terms of the momentum in a frame comoving with the
solar wind.  Assuming that the large-scale structure of the field
is stationary in a frame corotating with the sun, and that irregularities
in the field are stationary in the local solar wind frame, we derived
expressions for the rate of change of the momentum, $p'$, as a function
of the pitch angle (eq.~[7]), of $z$, the distance from the sun
along the magnetic field (eq.~[8]), and of $\mu'$, the pitch-angle cosine
(eq.~[10]).

We now incorporate these expressions into a Fokker-Planck equation for
the transport of energetic particles in the interplanetary magnetic field.
Our equation is a modified version of those developed by Jokipii (1966)
and Earl (1976), following the notation of Ng \& Wong (1979).  From now
on, we will always be working with $p$ and $\mu$ in the local solar wind
frame, and for convenience, these variables are now written without primes.
We start with a Fokker-Planck equation:
$$\eqalign{
{\pd F(t,\mu,z,p)\over\pd t} =
  -&{\pd\over\pd z}\left({\Delta z\over\Delta t} F\right)
  - {\pd\over\pd\mu}\left({\Delta\mu\over\Delta t} F\right)\cr
  +&{\pd\over\pd\mu}\left[{\varphi(\mu)\over 2} {\pd\over\pd\mu}
   \left({E'\over E}F\right)\right]
  -{\pd\over\pd p}\left({\Delta p\over\Delta t} F\right),\cr
}$$
where $\varphi(\mu)$ is the pitch-angle scattering coefficient, the
distribution function $F$ is given by
$$
F(t,\mu,z,p) = {d^3N\over dzd\mu dp},
$$
and $N$ represents the number of particles
inside a given flux tube (Ng \& Wong 1979).
Note that there is a factor of $E'/E=1-\mu v v_{\rm sw}\sec\psi/c^2$
in the pitch-angle scattering term, which relates our distribution
function to one defined in terms of time and position in the local
solar wind frame (Webb \& Gleeson 1979; see also the scattering terms
of Skilling 1975, Earl 1984).
Then replacing $\Delta z/\Delta t$, $\Delta\mu/\Delta t$, and
$\Delta p/\Delta t$ with $\dot z$, $\dot\mu$, and $\dot p$ from
equations (7), (8), and (10), we derive the following transport equation:
$$\eqalignno{
&{\pd F(t,\mu,z,p)\over\pd t} = -{\pd\over\pd z}\mu
  vF(t,\mu,z,p)\qquad\qquad\qquad\qquad\qquad\qquad\qquad
  &\rm (streaming) \cr
&\quad-{\pd\over\pd z}\left(1-\mu^2{v^2\over c^2}\right)
     v_{\rm sw}\sec\psi F(t,\mu,z,p) &\rm (convection) \cr
&\quad-{\pd\over\pd\mu}{v\over 2L(z)}
  \left[1+\mu{v_{\rm sw}\over v}\sec\psi
  -\mu{v_{\rm sw}v\over c^2}\sec\psi\right]&\cr
&\quad\quad\cdot(1-\mu^2)F(t,\mu,z,p)
  &{\rm (focusing)}\cr
&\quad+{\pd\over\pd\mu}v_{\rm sw}\left(\cos\psi{d\over
  dr}\sec\psi\right)\mu(1-\mu^2)&\cr
&\quad\quad\cdot F(t,\mu,z,p)
  &{\rm (differential\ convection)}\cr
&\quad+{\pd\over\pd\mu}{\varphi(\mu)\over 2}{\pd \over\pd\mu}
  \left(1-\mu{v_{\rm sw}v\over c^2}\sec\psi\right)F(t,\mu,z,p)
  &{\rm (scattering)}\cr
&\quad+{\pd\over\pd p}pv_{\rm sw}\left[{\sec\psi\over2L(z)}(1-\mu^2)
  +\cos\psi{d\over dr}\sec\psi\mu^2\right]&\cr
&\quad\quad\cdot F(t,\mu,z,p).
  &{\rm (deceleration)\rlap{\quad(11)}}\cr
}$$

The numerical code can accommodate any desired functional form for
the angle between the field line and the radial direction, $\psi(z)$,
the focusing length, $L(z)$, and the pitch-angle scattering coefficient,
$\varphi(\mu)$.  In this work, we use the Archimedean field model of Parker
(1958) with $b=0$, so that
$$\eqalign{
\cos\psi &= {R\over\sqrt{r^2+R^2}}\cr
L &= {r(r^2+R^2)^{3/2}\over R(r^2+2R^2)},\cr
}$$
where $R=v_{\rm sw}/\Omega\sin\theta$, $\Omega$ is the angular
rotation rate of the sun, and the radius is in turn a function of $z$.

We also use
$$\varphi (\mu) = A|\mu|^{q-1}(1-\mu^2),\eqno(12)$$
which was derived in the context of quasilinear scattering theory
(Jokipii 1971; Earl 1973), and is adopted here as a convenient
and commonly understood parameterization of the pitch-angle scattering.
Some care is required when using equation~(12),
because of the singularity at $\mu=0$.  In particular, when using a finite
difference method to numerically solve the transport equation, Ng \& Wong
(1979) used an effective scattering coefficient given (in our notation) by
$$\varphi_{\rm eff}(\mu) = {v\over 2L(z)}(1-\mu^2)
  {\Delta\mu\over \tanh\{{v\over 2AL(z)}
  {\scriptstyle\left[I(\mu+\Delta\mu/2)-I(\mu-\Delta\mu/2)\right]}
  \}},$$
where
$$I(\mu)\equiv {\rm sgn}(\mu){|\mu|^{2-q} \over 2-q}.$$
and $\Delta\mu$ is the grid spacing in the $\mu$ coordinate.
Although Ng and Wong only used this form for $\mu$ near zero, we use
it for all values of $\mu$, noting that
$\varphi_{\rm eff}(\mu)\rightarrow\varphi(\mu)$ when $\mu\gg\Delta\mu$.
By using $\varphi_{\rm eff}(\mu)$, we achieve a great improvement in
the accuracy of the numerical solutions.

Finally, it is convenient to describe the pitch-angle scattering
in terms of a spatial mean free path, $\lambda$, given by
$$\lambda \equiv {3D\over v},$$
where $D$ is a spatial diffusion coefficient.  Considering pitch-angle
scattering alone, one can derive a diffusion coefficient of
$$D = {v^2\over 4} \int_{-1}^1 {(1-\mu^2)^2\over \varphi(\mu)} d\mu$$
(Jokipii 1968, Hasselmann \& Wibberenz 1968).
If we replace the continuous variable $\mu$ with a discrete set of
grid points, the actual diffusion coefficient in the simulation is given by
$$D = {v^2\over 4} \sum_{\mu}{(1-\mu^2)^2\over \varphi_{\rm eff}(\mu)}
\Delta\mu,$$
where the sum is over $\mu$ values halfway between grid points,
and $\varphi_{\rm eff}(\mu)$ is calculated in the limit of no focusing.
We then find the scattering amplitude, $A$, that is needed
to generate the desired mean free path.

The transport equation presented here is different from those generally used
to model solar modulation (Jokipii 1992)
in that the distribution function depends on
the pitch angle, in addition to the position and momentum.
Therefore, this framework can be used to study anisotropic
distributions of galactic cosmic rays subject to solar modulation.

This transport equation can be shown to be equivalent to a
special case of the transport equation
derived by Skilling (1975), with the Alfv\'en speed set to zero and
the solar wind speed set to $v_{\rm sw}^c$.

The essential difference between the present derivation and that of
Luhmann (1976) is that the latter derivation was
performed in the fixed frame.  Scattering due to fluctuations moving at a
fixed speed was then expressed by four terms involving the diffusion
coefficients, $D_{\mu\mu}$, $D_{\mu p}=D_{p\mu}$, and $D_{pp}$ (see
also Schlickeiser 1989).  It is important to note that these four terms
mathematically conspire to act as the single scattering term in equation~(11)
(Earl 1984).  In fact, when viewed in the local solar wind frame, where the
magnetic fluctuations are fixed, scattering conserves the momentum,
so the $D_{pp}$ term of Luhmann (1976) only arises
upon transformation to the fixed frame (see Figure~1).
For a numerical solution, it would be difficult to evaluate the four
diffusion terms properly so that they would behave as the single term in
equation (11), and errors in the apparent $p$-diffusion would probably
dominate the true deceleration.

\bigbreak
\medskip
\centerline{3. \  NUMERICAL METHOD}
\medskip
The above transport equation was solved directly by means
of a finite-difference method.  The method was adapted from that
of Ruffolo (1991), but many modifications were needed to accommodate
adiabatic deceleration and other effects of the solar wind.
In particular, the distribution function must now be computed for
various values of the momentum, $p$.  In order to avoid a great
increase in the computer time required, the numerical code was
designed to provide accurate results even for widely spaced values
of $p$.  Thus while $F(t,\mu,z,p)$ is calculated for an
evenly spaced grid of $(t,\mu,z)$ coordinates, values of $p$ can be
chosen according to the user's convenience, e.g., corresponding to
momenta for which experimental data are reported.

When solving equation (11), we need to specify boundary conditions at
the edges of the $(t,\mu,z,p)$ domain.  As $\mu\rightarrow\pm 1$, we
require the $\mu$-flux,
$$\eqalignno{
S_\mu(t,\mu,z,p) = &\left\{{v\over 2L(z)}\left[1+\mu{v_{\rm sw}\over v}
  \sec\psi - \mu{v_{\rm sw}v\over c^2}\sec\psi\right](1-\mu^2)\right.&\cr
  &\quad\left. -v_{\rm sw}\left(\cos\psi{d\over dr}\sec\psi\right)\mu(1-\mu^2)
  \right\}F(t,\mu,z,p)&\cr
  &\quad-{\varphi(\mu)\over 2} {\pd\over \pd\mu}
  (1-\mu{v_{\rm sw}v\over c^2}\sec\psi) F(t,\mu,z,p),&(13)\cr
}$$
to vanish so that no particles ``flow'' to the non-physical regions where
$|\mu|>1$; this is satisfied by the terms including the factor
$(1-\mu^2)$, and by our choice for the scattering coefficient,
$\varphi(\mu)$.  Other boundary conditions can be adapted to the
problem at hand.  For simulations of the transport of particles
coming directly from a solar flare, the initial distribution function,
$F(t,\mu,z,p)|_{t=0}$ (where $t$ is set to zero at the time of the
flare), is only non-zero near $z=0$, i.e., near the
sun.  At the $z$-boundaries, we allow particles to flow out and set
the influx to zero, that is, particles are absorbed at the
boundaries.  In practice, the outer $z$-boundary can be set
sufficiently far from the sun that it has no effect on the
distribution of particles at the point of interest during the course of
the simulation, and the inner $z$-boundary near the Sun has only a small
effect, because intense focusing there reflects most incoming particles
before they reach the boundary (Ng \& Wong 1981).
Finally, the distribution function vs.~momentum at a fixed $s=vt$ is assumed
to fall off as $p^{-\delta}$ above the highest simulated value of $p$
(see discussion below).  This is equivalent to fixing the $p$-flux at
the upper $p$-boundary.

We solve the transport equation by applying a technique known as operator
splitting, in which $F(t,\mu,z,p)$ is sequentially updated according to
individual terms or groups of terms on the right-hand side of equation (11).
The procedure can be physically interpreted as having particles
alternately undergo changes in $\mu$, $p$, and $z$ during each time step.
As the time step is shortened, this sequence provides a more accurate
approximation to the simultaneous action of these processes.
The resulting procedure for updating $F(t,\mu,z,p)$ from $t$ to $t+\Delta t$
is as follows (in equations with a left-pointing arrow, the terms on the
left hand side refer to updated values):

{\parskip=20pt plus 0.5pt minus 0.5 pt

\itemitem{1.} Update the distribution function
with half the effect of pitch-angle scattering, focusing, and
differential convection, i.e., processes which affect $\mu$.
Step~4 accounts for the second half; the treatment of these processes
both before and after steps 2 and 3 improves the convergence of the method.
For fixed values of $z$ and $p$, the $\mu$-flux, $S_\mu$, is evaluated
between each pair of neighboring grid points, $\mu$ and $\mu+\Delta\mu$.
Using equation (13) and the finite difference approximation for
derivatives, we obtain
$$\eqalign{
S_\mu(\bar\mu)\approx  &\left\{{v\over 2L(z)}
    \left[1+\bar\mu{v_{\rm sw}\over v}\sec\psi
    -\bar\mu{v_{\rm sw}v\over c^2}\sec\psi\right]\right.\cr
  & \left.-v_{\rm sw}\left(\cos\psi{d\over
    dr}\sec\psi\right)\bar\mu\right\}\cr
  &\quad\cdot(1-\bar\mu^2)\left[F(\mu+\Delta\mu)+F(\mu)\over 2\right]\cr
  &-{\varphi_{\rm eff}(\bar\mu)\over 2}
  \left[F'(\mu+\Delta\mu)-F'(\mu)\over \Delta\mu\right],\cr
}$$
where $\bar\mu\equiv\mu+\Delta\mu/2$ and $F'=(E'/E)F=(1-\mu
vv_{\rm sw}\sec\psi/c^2)F$.  The $\mu$-flux can be calculated using
$f$ at the start of the step (explicitly), at the end of the step (implicitly),
or alternately explicitly and implicitly for better stability and accuracy
(Crank \& Nicolson 1947).  Using the last
approach, the code first solves the coupled explicit equations,
$$\eqalignno{\noalign{\vskip6pt}
F(\mu) &\leftarrow F(\mu)
-\left(\Delta t\over 4n\right)\left[{S_\mu(\mu+\Delta\mu/2)
-S_\mu(\mu-\Delta\mu/2) \over \Delta\mu}\right],\cr
\noalign{\vskip6pt}}$$
for each grid point, $\mu$, then solves the implicit equations,
$$\eqalignno{\noalign{\vskip6pt}
F(\mu)+\left(\Delta t\over 4n\right)\left[{S_\mu(\mu+\Delta\mu/2)
-S_\mu(\mu-\Delta\mu/2) \over \Delta\mu}\right]
&\leftarrow F(t,\mu),\cr
\noalign{\vskip6pt}}$$
and repeats the process for a total of $n$ explicit and $n$ implicit
substeps.  Each substep represents a time increment of
$\Delta t/(4n)$, so these $2n$ steps account for scattering
and focusing over half of $\Delta t$.  The number, $n$, is repeatedly doubled
until the resulting flux changes by less than the preset absolute tolerance
or the preset relative tolerance.

\itemitem{2.} Update the distribution function with the effect
of deceleration, i.e., the systematic decrease in $p$.
This step requires particular care, so that numerical errors do not
dwarf the small effect of deceleration.
Since the spacing between $p$-grid points is typically much larger
than the change in momentum during a single time step, it is necessary
to interpolate the distribution function between $p$-grid points.

\itemitem{}To minimize the error of this approximation, the
interpolation should be between $p$-grid points at the same time, $t$, or
at the same ``distance traveled,'' $s=vt$, depending on the physical
situation.  For example, consider the distribution of particles
after a solar flare.  For two neighboring values of the momentum,
$p_1$ and $p_2$, where $p_1<p_2$, the $\mu$-$z$ distributions will
evolve differently vs.~$t$, because the particles at $p_2$ move
faster than those at $p_1$.  However, their evolution vs.~$s=vt$ is
quite similar.  Thus geometric interpolation at a constant $s$ provides an
accurate estimate of the distribution function at intermediate $p$ values.
In contrast, interpolation at a constant $t$ is inaccurate, and can
also introduce an erroneous, non-zero flux of particles well beyond a distance
$vt$ from the sun.  Therefore, interpolation at constant $s$ is more
appropriate for modeling the transport of solar flare particles;
interpolation at constant $t$ would be more appropriate for certain
other astrophysical situations.

\itemitem{}The transport equation for deceleration alone can be written as
$$
{\pd \over \pd t}F(t,\mu,z,p) = {1\over\tau_{\rm d}}{\pd\over\pd
p}pF(t,\mu,z,p),
$$
where the deceleration time, $\tau_{\rm d}$, is given by
$$
{1\over\tau_{\rm d}} = v_{\rm sw}\left[{\sec\psi\over2L(z)}(1-\mu^2)
  +\cos\psi{d\over dr}\sec\psi\mu^2\right].\eqno(14)
$$
This in turn gives us
$$
{\pd \over \pd t}pF(t,\mu,z,p) = {1\over\tau_{\rm d}}{\pd\over\pd
\ln(p/p_0)}pF(t,\mu,z,p),
$$
for a fixed reference momentum $p_0$, which has the solution
$$
pF(t+\Delta t,\mu,z,p) = pe^{\Delta t/\tau_{\rm d}}F(t,\mu,z,pe^{\Delta
t/\tau_{\rm d}}). \eqno(15)
$$
When the interpolation is performed at a constant $t$, this
formula can be used directly.  When interpolating at a constant $s$,
we note that the product $pF$ is constant along characteristics,
which are straight lines of slope $-1/\tau_{\rm d}$ on a semi-logarithmic
graph of $p$ vs.\ $t$ (see Fig.~4).
We then find the intersection between the curve of constant $s$
passing through the grid point $(t_i,p_j)$ and the characteristic
passing through $(t_{i+1},p_j)$, where $t_{i+1}=t_i+\Delta t$.  At this
characteristic-intersection point, $(t^{*},p^{*})$, the distribution
function is calculated by geometric interpolation between $F(p_j)$ and
$F(p_{j+1})$ at the next higher $p$-grid point (or for the highest $p$-grid
point, by assuming that $F\propto p^{-\delta}$ for a constant $\delta$ lines
of constant $s$).  We then use
$$
pF(t_{i+1},\mu,z,p_j) = p^{*}F(t^*,\mu,z,p^{*})
$$
to calculate the new value of the distribution function.

\itemitem{}Note that we are free to choose a different value of $\Delta t$
for each value of $p$.  We therefore set $\Delta t = \Delta s/v$ for a
fixed value of $\Delta s$, so that $F$ is updated to $s_{i+1}=s_i+\Delta s$
for each $p$-grid point (see Fig.~4).

\itemitem{}This characteristic-intersection method fails when the
logarithmic slope of the characteristics, $-1/\tau_d$, is steeper than that
of the constant-$s$ curve, $-\gamma^2 / t$, or when
$$
t > \gamma^2\tau_{\rm d}.
$$
The deceleration time, $\tau_{\rm d}$, is at least 5.7 days near the Earth,
for an Archimedean field and a solar wind speed of 400 km/s.
Thus this code is valid for simulating ``well-connected'' solar flare
events, for which the flare and detector are on nearby magnetic
field lines, and the flare event has a short duration;
for poorly-connected events, the code would need to be modified to
switch to interpolation at constant $t$ for later times, and also to
take into account drifts and diffusion perpendicular to the field.

\itemitem{3.} Account for streaming and convection in the $z$-direction.
The streaming term alone can be treated (Ruffolo 1991) by simply moving
the distribution function from one $z$-grid point to another:
$$F(i\Delta\mu,z,p)  \leftarrow  F(i\Delta\mu,z - i\Delta z,p),$$
where $i$ labels the $\mu$-grid points, $\mu=i\Delta\mu$, and
$\Delta z = \Delta\mu(v\Delta t)$.  This is an exact solution to a transport
equation for streaming alone,
$$
{\pd F(t,\mu,z,p) \over \pd t} = -\mu v{\pd \over \pd z} F(t,\mu,z,p),
$$
and since no finite difference approximation is used, there is
no numerical ``diffusion''.  When convection is included, this
approach can be modified by occasionally moving the distribution
function forward by an additional step, $\Delta z$.  Thus
numerical diffusion is still absent, i.e., a sharp pulse or gradient
in the distribution function can be maintained intact.  There is
a price to pay for the elimination of numerical diffusion:
because the convection speed, $(1-\mu^2{v^2/c^2})v_{\rm
sw}\sec\psi(z)$, is not constant with respect to $z$, ``holes'' where $F=0$
can appear in the $\mu$-$z$ plane.  Fortunately, these holes are
rapidly filled in by pitch-angle scattering (in steps 1 and 4),
and on average the distribution is rarefied in the proper
manner.  For this reason, the results presented here are averaged
over a set of nearby $z$-grid points.

\itemitem{4.} Repeat step 1, updating the pitch-angle distribution for
the second half of the scattering, focusing, and differential
convection effects.  The result of this step is $F(t+\Delta t,\mu,z,p)$.

Starting with the initial condition, $F(0,\mu,z,p)$,
we repeat this procedure for each time step, $\Delta t$, to
generate the particle distribution function for all times of interest.

} 

\bigbreak
\medskip
\centerline{4. \  NUMERICAL SIMULATIONS}
\medskip
Several simulations were performed to test the accuracy of the new
computer code.  For example, when $F\propto p^{-\delta}$ and only
the deceleration step is enabled, an
exponential decay in the flux was observed, with the proper time
constant as given by equation~(14).  Simulation results for $v_{\rm
sw}=0$ matched those of the code used in Ruffolo (1991), in which
solar wind effects were not included.  That code, in turn,
successfully fit the observed intensity and pitch-angle distribution
vs.\ time for neutron-decay protons from the solar flare events of
3 June 1982 and 25 April 1984.  It has also been tested against the codes
of Ng \& Wong (1979), Bieber et al.~(1980),
and more recently, against those of Earl (1987) and Pauls \& Burger
(1993), as described by Earl et al.~(1994).

The remainder of this section describes sample simulations of
the transport of solar flare protons.  As described earlier,
the momentum, $p$, and the pitch-angle cosine, $\mu$, are defined in
the local solar wind frame; quantities in the fixed frame can
be derived by a Compton-Getting transformation.  The simulations were
performed for five momentum values corresponding to kinetic energies of 2, 6,
20, 60, and 200 MeV.  The initial condition is that protons are
concentrated near the sun.  For demonstration purposes, the initial
distribution is proportional to $p^{-5}$; this is somewhat steep, but
not unusual for solar flare protons (see, e.g., Evenson et al.~1984).
The initial distribution is concentrated at the highest $\mu$-grid
point, because strong focusing near
the sun herds the particles into the forward direction, and
the normalization is such that the integral of $F$ over $\mu$ and $z$
for a kinetic energy of 2 MeV is equal to one.  For other parameters,
typical values were chosen: the solar wind speed is 400 km/s,
and the scattering parameter $q$ and the mean free path $\lambda$ are
1.5 and 0.3 AU, respectively, for all values of $p$, $z$, and $t$.
The step size, $\Delta s=v\Delta t$, is 0.005 AU up to $s=0.5$ AU, 0.01 AU
up to $s=1$ AU, etc.  The $\mu$-grid spacing, $\Delta\mu$, is 2/25
(i.e., for 25 values of $\mu$), and $\Delta z$ is set equal to
$\Delta s\Delta\mu$ in order to accurately evaluate the streaming
term (see \S 3), i.e., $\Delta z$ is $4\times10^{-4}$ AU up to
$s=0.5$ AU, $8\times10^{-4}$ AU up to $s=1$ AU, etc.  A typical simulation
required approximately 2.5 hours of CPU time on an IBM RISC/6000 computer.

The distribution of 2 MeV protons vs.~$\mu$ and $z$ is shown in
Fig.~5 for different values of $s=vt$, the distance traveled by the
protons since the occurrence of the flare, or their maximum possible distance
from the sun in the absence of convection.  Panels a-d show results
for $s$ values of 0.5, 1, 2, and 4 AU, respectively.  The general
features are typical of focused transport.  The protons are initially
concentrated in a coherent pulse (Earl 1976)
over a narrow range of $z$ values.  The distribution is strongest
near $\mu=1$, for which particles move directly along the magnetic
field and away from the sun.  The distribution function decreases sharply at
$\mu=0$ (where the pitch angle is 90$^\circ$), because the scattering
coefficient approaches zero as $\mu\rightarrow 0$, which inhibits scattering
to $\mu<0$ (see eq.~[12]).

Note that for each of the plots in Fig.~5,
the right boundary is at $z=s=vt$.  Since this is the maximum
possible distance from the sun, in the absence of convection,
the protons would reach the right edge of the plots if
they traveled directly along the field, with $\mu=1$.
In fact, the pulse moves somewhat slower than $v$ because the protons
are distributed over a range of $\mu$ values.
Furthermore, a comparison of panels a-c shows that the pulse moves
progressively more slowly.  This is because the effect of focusing,
which is proportional to the logarithmic gradient of $B$, becomes weaker
as the pulse moves farther from the sun, allowing the average $\mu$
to decrease.  Later on, when $s=1.0$ (panel c), much of the pulse has
spread to $\mu<0$; from this point on, the distribution is
dominated by a nearly Gaussian ``wake'' (Earl 1976) that gradually
spreads in $z$ as the coherent pulse fades away.

To isolate the effect of the solar wind,
simulations were also performed with the solar wind velocity set to
zero.  The difference, $\Delta F$, between $F$ for $v_{\rm sw}=400$ km/s
and $F$ for $v_{\rm sw}=0$ is displayed in panels e-h of Fig.~5.
In each case, the difference is negative behind the pulse,
and positive at the leading edge.  This can be explained by
the effects of solar wind convection and adiabatic deceleration.
For small $s$, convection displaces the narrow pulse by a distance
comparable to its width, yielding the bipolar distribution of panel~e.
Later on, deceleration has a greater cumulative effect, causing a
systematic decrease in $F$ (see panel~h).  Note that each
plot of $\Delta F$ is drawn to the same scale as the corresponding
plot of $F$, so for this case the solar wind has a substantial impact
on the cosmic ray distribution.

The simulated intensity vs.\ the distance traveled, $s$, is shown in
Fig.~6, both with and without the effect of the solar wind.  The
intensity shown here is the directional average of the distribution
function, $F$, for fixed values of $z$ and $p$.
The basic features of a rapid rise and a slow decay of the intensity
are characteristic of observed solar cosmic-ray distributions after a flare
event (in the absence of interplanetary shock effects) and are in
qualitative agreement with the results of the older, diffusive
propagation model (e.g., Wibberenz et al.~1989).
Since we have assumed that the scattering mean free path, $\lambda$, is
independent of energy, the evolution of the intensity vs.\ $s$ is
also energy independent in the absence of solar wind effects.
Thus the points for $v_{\rm sw}=0$ (pluses) for different
energies differ by a constant logarithmic offset, and any
energy dependence of the intensity evolution for $v_{\rm sw}=400$
km/s (solid circles) is due to the effect of the solar wind.
For kinetic energies of 60 and 200 MeV, the two sets of points are nearly
coincident, indicating that the solar wind has only a small effect on the
transport of higher-energy solar cosmic rays; this is because the
solar-wind dependent terms in $\pd F/\pd s$ depend on the ratio
$v_{\rm sw}/v$, which is small for higher energy particles.  In contrast,
there is a significant difference for the lower energies, mainly due to the
effect of solar wind convection at early times and of adiabatic deceleration
at later times.

One possible diagnostic of the evolution of the
intensity of solar cosmic rays is its decay at late times, which
typically appears as a straight line on a semi-logarithmic plot.
The slope can then be related to $\lambda$, the mean free path of
interplanetary scattering, based on a model of the interplanetary transport.
Our results show that such models should take adiabatic deceleration
into account for kinetic energies on the order of 20 MeV or below,
or else the derived value of $\lambda$ will be in error.  To quantify this
error, let the ``apparent'' mean free path, $\lambda_a$, be defined as the
value needed for a simulation with no solar wind to match the late-time decay
slope of a simulation for the true $\lambda$ that includes solar wind
effects.  Table~1 shows values of $\lambda_a$, where the true $\lambda$
is 0.3 AU.  Results are shown both for a spectral index, $\delta$, of
5 as before, and for a more typical case of $\delta=2.5$.
Note that for three cases (including 2 and 6 MeV protons shown in
Fig.~6), no reasonable value of $\lambda_a$ can generate such a steep slope;
for $\lambda_a > 0.5$ AU, the decay phase starts after $s=4$ AU.
In determining $\lambda$ from observational data, it is possible to
achieve an accuracy as good as 20\% (Ruffolo 1991), but Table~1
shows that for most cases, the error in neglecting deceleration would
be greater than this.  Even for the rather high energy of 20 MeV,
the mean free path would be overestimated by 63\% for a flare with
a spectral index of 5.

Returning to other results,
simulated pitch angle distributions of 2 MeV protons are shown in
Fig.~7 for selected $s$ values.  For both $r=0.3$ AU and $r=1.0$
AU, the pitch-angle distributions before or at the time of the peak
flux are highly anisotropic in the forward direction (outward along
the magnetic field, or to the right in Fig.~7), which reflects the
highly collimated nature of the coherent pulse.  At the other
extreme, the distributions for late times are nearly isotropic.  For
intermediate values of $s$, on the trailing edge of the pulse,
more unusual shapes are found, as shown here for
$s=0.6$ AU and $s=2.52$ AU.  Since the trailing edge comprises
protons traveling slightly slower than those at the pulse peak, the
intensity there is highest at a $\mu$ value intermediate between 0 and 1.
Note also the sharp gradient at $\mu=0$, due to the drop in the scattering
coefficient as discussed earlier.  More examples of such
distributions can be found in Bieber et al.~(1980).

Finally, we have performed simulations which examine the effects of
individual, solar-wind dependent terms in the transport equation.
These include terms affecting $\dot p$ (deceleration), $\dot z$
(convection), and $\dot\mu$ (differential convection and corrections
to focusing).  We ran simulations where one of these factors was
included, and the other two were disabled.  The terms affecting
$\dot\mu$ were found to have a very small effect, so we
concentrate on the deceleration and convection terms.  Fig.~8 shows
results for $\delta=5$, $v_{\rm sw}=400$ km/s, $T=2$ MeV, and $r=1$ AU,
with 1) no solar wind effects, 2) convection only, 3) deceleration
only, and 4) all solar wind effects.  The main effect of convection
is to speed up the pulse, so that protons start to arrive sooner.
At later times, convection leads to a slightly steeper decay slope,
because the peak in the distribution is convected farther from the
point of observation, and because of the rarefaction due to the
increase in the convection speed with increasing $z$.  Deceleration
creates a widening deficit in the intensity vs.~$s$, and accounts for
most of the change in the decay slope upon the inclusion of solar wind
effects.  Finally, the simulation including all effects exhibits both an
earlier rise and a steeper decay at late times.

Fig.~8 seems to indicate an additive relationship between the slope
changes due to different effects, which has been employed by previous
authors (e.g., Murray et al.~1971).  Combining all significant effects,
we can express the rate of decay of the intensity vs.~$s$ as
$$
{1\over vT} \approx {1\over vT_{\rm ssf}} + {1\over vT_{\rm d}}
 + {1\over vT_{\rm c}},
$$
where $T_{\rm ssf}$ is the exponential-decay time constant for streaming,
scattering, and focusing, i.e., in the absence of solar wind effects,
$T_{\rm d}$ is that due to including deceleration,
and $T_{\rm c}$ is that due to including solar wind convection.
These four terms, as derived from the simulations shown in Fig.~8, adding one
process at a time, are 0.2480, 0.1327, 0.0871, and
0.0345 AU$^{-1}$, respectively, so the above approximation
is in error by 0.0063 AU$^{-1}$ (due to interactions between the effects of
deceleration and convection).  The derived intensity-decay times are then
$T=0.3572$ days, $T_{\rm ssf} = 0.6676$ days, $T_{\rm d} = 1.017$ days, and
$T_{\rm c} = 2.57$ days.

It is instructive to compare the above decay time for deceleration
with estimates based on a hypothetical transport equation with
deceleration alone.  From equation~(15), we find that the
momentum-decay rate, $\tau_{\rm d}$, leads to an
a intensity-decay time of $\tau_{\rm d}/(\delta-1)$ due to the advection
of $F$ from higher momenta.  For a nearly isotropic pitch-angle
distribution at late times, we expect that
$$
{1\over\tau_{\rm d}} = {2\over 3} {v_{\rm sw}\over r},
$$
and for $r=1.0$ AU, $\tau_{\rm d} = 6.49$ days.  The effective value of
$\delta$ is actually 4.65, due to the enhanced decay of the
lower-energy flux.  Thus we might expect that
the intensity-decay time due to including deceleration would be
$T_{\rm d} = 1.78$ days; however, this is 75\% longer than the value
observed from the simulations.

A careful examination of the operation of the code shows that step 2
(see \S 3) generates a decay in the intensity at the expected rate.
However, the other steps produce a faster decay when deceleration is
included, because of the cumulative effect on the shape of the
distribution function.  During this late-time, diffusive r\'egime,
the processes of streaming and scattering behave as a spatial
diffusion process, yielding a diffusive decay rate proportional to
$\pd^2 F/\pd z^2$.  Because the rate of deceleration is spatially non-uniform,
it turns out that deceleration for $s<4.0$ AU increases the magnitude of the
diffusive decay rate in the neighborhood of $r=1.0$ AU.  By this means, the
influence of deceleration on the intensity decay is effectively enhanced by
75\% over that due to momentum advection alone.

\bigbreak
\medskip
\centerline{5. \  DISCUSSION AND CONCLUSIONS}
\medskip
The results of the preceding section justify the consideration of
adiabatic deceleration in focused transport models.  One might
expect that the effect of deceleration could be estimated
from equation~(15), and that this estimated decay rate
could be simply added to that due to the other processes (e.g., Murray
et al., 1971).  However, we have shown that the effect of including
deceleration can be much greater than the simple estimate, because the
effect is greatly amplified by an interaction with diffusive processes.
The magnitude of amplification depends on the cumulative effect of
deceleration on the shape of the distribution function, $F$, which in turn
depends on the detailed history of the distribution function.  Thus to
accurately evaluate the effect of deceleration on the transport of solar
cosmic rays, it is necessary to fully incorporate the process into a
numerical code for focused transport, as we have done here.

We have also pointed out possible errors due to neglecting
deceleration in the determination of the mean
free path, $\lambda$, from the late-time intensity decay rate, for various
kinetic energies and two values of the spectral index.  This
naturally leads to the question of how $\lambda$ would be affected by
other processes which have been neglected.  The most important of
these are diffusion and drifts perpendicular to the magnetic field.
A very rough estimate shows that these effects are of the same order of
magnitude as adiabatic deceleration, so these are important
processes which should be taken into account.  [Note, however, the
justification for their neglect in Ng (1987).]  Efforts in this direction
have been hampered by unresolved issues, such as the magnitude of the
diffusion coefficient, whether perpendicular
spreading occurs more in the interplanetary medium (as diffusion) or
in the corona, over what distance or speed coronal
propagation takes place, etc.  In certain flare events, a strong
interplanetary shock is generated, and there are various mechanisms
by which particles can be accelerated or their transport can be
affected [see Jones (1990) for a relevant transport equation].
Other conceivable effects on the particles, such as centripetal
or Coriolis acceleration in the corotating frame, can be shown to
be orders of magnitude smaller.

In addition to the late-time decay slope, one can also make use of
other aspects of the simulation results to fit observed intensity and
pitch-angle distributions.  The simulation results could
serve as templates or response functions for an impulsive injection
of particles, so that the time dependence of the injection can be
adjusted to fit the observations (see Ma Sung \& Earl 1978; since
perpendicular diffusion in the interplanetary medium is neglected, this is
tantamount to assuming that all perpendicular motion takes place in the solar
corona).  Note that the effect of convection is also important
(see Figure~8), especially in regard to the timing of the
pulse's arrival, which provides the most sensitive information about
the time dependence of the injection (Ruffolo 1991).

Other applications of this numerical method can be envisioned.  It
could be applied to various situations in which adiabatic
deceleration or solar wind affects are known to play a major role in cosmic
ray transport, such as Forbush decreases, shock acceleration, or solar
modulation.  Since the key feature of this work is the simultaneous treatment
of transport in position, momentum, and pitch angle,
the method would most useful when there is a significant anisotropy
in the directional distribution; otherwise, existing numerical or
analytic methods would suffice.

In summary, we have re-examined the first-order effects of the solar
wind on cosmic-ray transport, including convection, deceleration, and
pitch-angle effects, from the point of view of focused transport
theory.  These effects are then incorporated into a numerical code
for simulating the distribution of energetic particles as a function
of time, pitch angle, position, and momentum.  This is used to
model the transport of solar flare protons, and it is found that
both convection and deceleration have a significant impact on the
evolution of the distribution.  In particular, the influence of
deceleration is greatly enhanced by interactions with the other processes.
In addition, this numerical method is expected to be applicable to a
variety of situations where solar wind and pitch-angle dependent effects
must be evaluated simultaneously.

\medskip

The author would like to thank Steve Arendt, John Bieber, Jim Earl,
Paul Evenson, Peter Meyer, and Eugene Parker
for valuable discussions regarding this work.  Special thanks go to
Chee Ng for his detailed comments as a referee for this article.  I am
grateful to Burin Asavapibhop, Pisit Lila\-pattana, Wantana Songprakob, and
Montien Tienpratip for programming work.  Finally, I would like to
thank the Laboratory for Astrophysics and Space Research at the University
of Chicago for kindly allowing remote access to their computing facilities.

\bigskip
\bigskip
\centerline{REFERENCES}

\parskip=0pt
\def\ref{\hangindent=20pt\hangafter=1\noindent}

\ref
Bieber, J.~W.~1977,
Ph.D.~thesis, Univ.~Maryland

\ref
Bieber, J.~W., Earl, J.~A., Green, G., Kunow, H., M\"uller-Mellin, R.,
\& Wibberenz, G.~1980,
J.~Geophys.~Res., 85, 2313

\ref
Crank, J., \& Nicolson, P.~1947,
Proc.\ Cambridge Phil.\ Soc., 43, 50

\ref
Compton, A.~H., \& Getting, I.~A.~1935,
Phys.~Rev., 2d ser., 47, 817

\ref
Dorman, L.~I.~1965,
Proc.~9th Internat.~Cosmic Ray Conf., 1, 292

\ref
Earl, J.~A.~1973,
ApJ, 180, 227

\ref
\underbar{\ \ \ \ \ \ \ }.~1976,
ApJ, 205, 900

\ref
\underbar{\ \ \ \ \ \ \ }.~1984,
ApJ, 278, 825

\ref
\underbar{\ \ \ \ \ \ \ }.~1987,
Proc.~20th Internat.~Cosmic Ray Conf., 3, 198

\ref
Earl, J.~A., Bieber, J.~W., Pauls, H.~L., \& Ruffolo, D.~1994
(in preparation)

\ref
Evenson, P., Meyer, P., Yanagita, S., \& Forrest, D.~J.~1984,
ApJ, 283, 439

\ref
Forman, M.~A.~1970,
Planet.~Space Sci., 18, 25

\ref
Gleeson, L.~J., \& Axford, W.~I.~1967,
ApJ, 149, L115

\ref
Hasselmann, K., \& Wibberenz, G.~1968,
Zs.~Geophysik, 34, 353

\ref
Jokipii, J.~R.~1966,
ApJ, 146, 480

\ref
\underbar{\ \ \ \ \ \ \ }.~1968,
ApJ, 152, 671

\ref
\underbar{\ \ \ \ \ \ \ }.~1971,
Rev.~Geophys.~Space Phys., 9, 27

\ref
\underbar{\ \ \ \ \ \ \ }.~1992,
in Astronomy and Astrophysics Encyclopedia, ed.\ S.\ P.\ Maran
(New York: Van Nostrand Reinhold; Cambridge: Cambridge Univ.\ Press), 141

\ref
Jokipii, J.~R., \& Parker, E.~N.~1970,
ApJ, 160, 735

\ref
Jones, F.~C.~1990,
ApJ, 361, 162

\ref
K\'ota, J., Mer\'enyi, E., Jokipii, J.~R., Kopriva, D.~A., Gombosi,
T.~I., \& Owens, A.~J.~1982,
ApJ, 254, 398

\ref
Luhmann, J.~G.~1976,
J.\ Geophys.\ Res., 81, 2089

\ref
Ma Sung, L.~S., \& Earl, J.~A.~1978,
ApJ, 222, 1080

\ref
Murray, S.~S., Stone, E.~C., \& Vogt, R.~E.~1971,
Phys.~Rev.~Lett., 26, 663

\ref
Ng, C.~K., \& Wong, K.-Y.~1979,
Proc.~16th Internat.~Cosmic Ray Conf., 5, 252

\ref
\underbar{\ \ \ \ \ \ \ }.~1981,
Geophys.~Res.~Lett., 8, 113

\ref
Ng, C.~K.~1987,
Solar Phys., 114, 165

\ref
Parker, E.~N.~1958,
ApJ, 128, 664

\ref
\underbar{\ \ \ \ \ \ \ }.~1965,
Planet.~Space Sci., 13, 9

\ref
Pauls, L.~H., \& Burger, R.~A.~1993,
Proc.~23rd Internat.~Cosmic Ray Conf., 3, 191

\ref
Roelof, E.~C.~1969,
in Lectures in High Energy Astrophysics, ed.\ H.\ Ogelman \& J.\ R.\
Wayland (NASA SP-199) (Washington: NASA), 111

\ref
Ruffolo, D.~1991,
ApJ, 382, 688

\ref
Schlickeiser, R.~1989,
ApJ, 336, 243

\ref
Singer, S.~F.~1958,
Nuovo Cimento, Ser.~10, 8, Suppl.~2, 334

\ref
Singer, S.~F., Laster, H., \& Lenchek, A.~M.~1962,
J.~Phys.~Soc.~Japan, 17, Suppl.~A-2, 583

\ref
Skilling, J.~1971,
ApJ, 170, 265

\ref
\underbar{\ \ \ \ \ \ \ }.~1975,
MNRAS, 172, 557

\ref
Webb, G.~M., \& Gleeson, L.~J.~1979,
Ap\&SS, 60, 335

\ref
Wibberenz, G., Kecskem\'ety, K., Kunow, H., Somogyi, A., Iwers, B.,
Logachev, Yu.~I., \& Stolpovskii, V.~G.~1989,
Solar Phys., 124, 353

\vfill\eject
\centerline{}
\vfill

\noindent Table 1. \ Apparent Mean Free Paths

$$\vbox{\halign{
\hfil#\hfil&\ \hfil#\hfil&\ \hfil#\hfil\cr
$T$ (MeV) & $\lambda_a$ (AU) & $\lambda_a$ (AU) \cr
          & for $\delta=2.5$ & for $\delta=5$   \cr
\noalign{\vskip10pt\hrule\vskip10pt}
2   & \dots$^a$ & \dots$^a$ \cr
6   & 0.47      & \dots$^a$ \cr
20  & 0.38      & 0.49      \cr
60  & 0.36      & 0.40      \cr
200 & 0.34      & 0.37      \cr
\noalign{\vskip10pt\hrule\vskip10pt}
\noalign{\noindent
$^a\,$Change in slope is too great to be}
\noalign{\noindent
fit by a reasonable value of $\lambda_a$.}
}}$$

\vfill\eject
\centerline{}
\medskip
\centerline{Figure Captions}
\medskip

\itemitem{Fig.~1.\quad} Illustration of the actions of adiabatic
focusing and pitch-angle scattering in a) the fixed frame and
b) the solar wind frame.  Focusing acts to increase $v_\parallel$
along lines of constant velocity in the fixed frame,
while scattering acts in either direction along lines of constant
velocity in the solar wind frame.  ``S'' indicates the velocity of
the solar wind frame relative to the fixed frame,
and ``F'' indicates the velocity of
the fixed frame relative to the solar wind frame.

\itemitem{Fig.~2.\quad} Schematic trajectory of a cosmic ray
particle alternately undergoing scattering and focusing, as viewed in
the solar wind frame.  The combination of the two processes leads to a
systematic deceleration of the particle.

\itemitem{Fig.~3.\quad} Schematic of the average magnetic field
({\it curved lines}) in the inner solar
system, illustrating the definitions of the variables
$z$ and $\psi$ for the point of interest ({\it solid circle near the top
of the figure}).

\itemitem{Fig.~4.\quad} Illustration of the
characteristic-intersection method for computing the effect of
deceleration.  The quantity $pF$ is constant along characteristics
({\it dotted lines}), which are straight lines of constant slope in this
semi-logarithmic plot.  Solid lines are for constant values of
$s=vt$.  The value of $F(t_{i+1},p_j)$ is derived from that
at the point of intersection, $F(t^*,p^*)$, which is in turn derived
by geometric interpolation between $F(p_j)$ and $F(p_{j+1})$ along a line of
constant $s$.

\itemitem{Fig.~5.\quad} {Panels a-d}: Distribution, $F$, of 2 MeV protons
vs.~the distance along the magnetic field, $z$, and the pitch-angle
cosine, $\mu$, in the solar wind frame
for $v_{\rm sw}=400$ km/s and {\it a)} $s=0.5$ AU, {\it b)} $s=1$
AU, {\it c)} $s=2$ AU, and {\it d)} $s=4$ AU, where $s=vt$ is the distance
traveled.  {it Panels e-h}: Difference, $\Delta F$, between
distributions for $v_{\rm sw}=400$ km/s and $v_{\rm sw}=0$ km/s,
showing the effects of the solar wind on the protons' transport for
{\it e)} $s=0.5$ AU, {\it f)} $s=1$ AU, {\it g)} $s=2$ AU, and
{\it h)} $s=4$ AU.  Note that adjacent panels are plotted to the same scale.

\itemitem{Fig.~6.\quad} Simulated intensity vs.~distance traveled,
$s$, for radii of 0.3 and 1 AU, kinetic energies of 2, 6, 20, 60, and
200 MeV.  Results are shown for solar wind velocities of 0 km/s ($+$)
and 400 km/s ($\bullet$).  Small irregularities in the latter are an
artifact of the numerical treatment of solar wind convection (see \S
3).  Squares indicate points for which pitch
angle distributions are shown in Fig.~7.

\itemitem{Fig.~7.\quad} Polar plots of the 2 MeV proton intensity
vs.~pitch angle in the solar wind frame
for various values of $r$ and $s$.  Each plot is
normalized to the same radius for a pitch angle of zero, which is toward
the right in each plot.

\itemitem{Fig.~8.\quad} The logarithm of the intensity
of 2 MeV protons vs.~distance traveled, $s$, for simulations that
included no solar wind effects ($+$), convection only ($\circ$),
deceleration only ($\times$), and all solar wind effects ($\bullet$),
for a radius of 1 AU.  Note that convection results in an earlier
arrival of protons, and deceleration causes a lower intensity and a
faster decay after the peak.

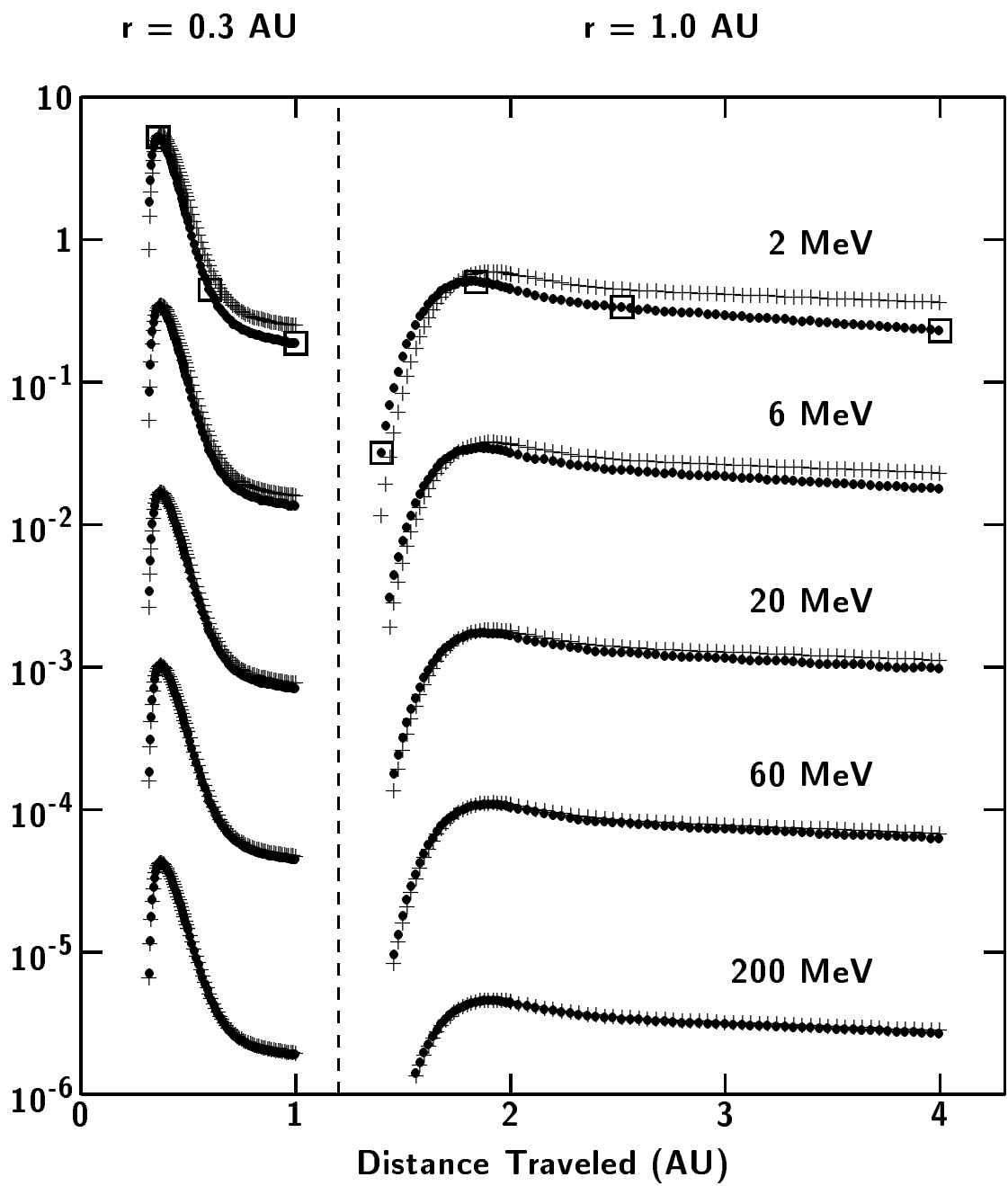
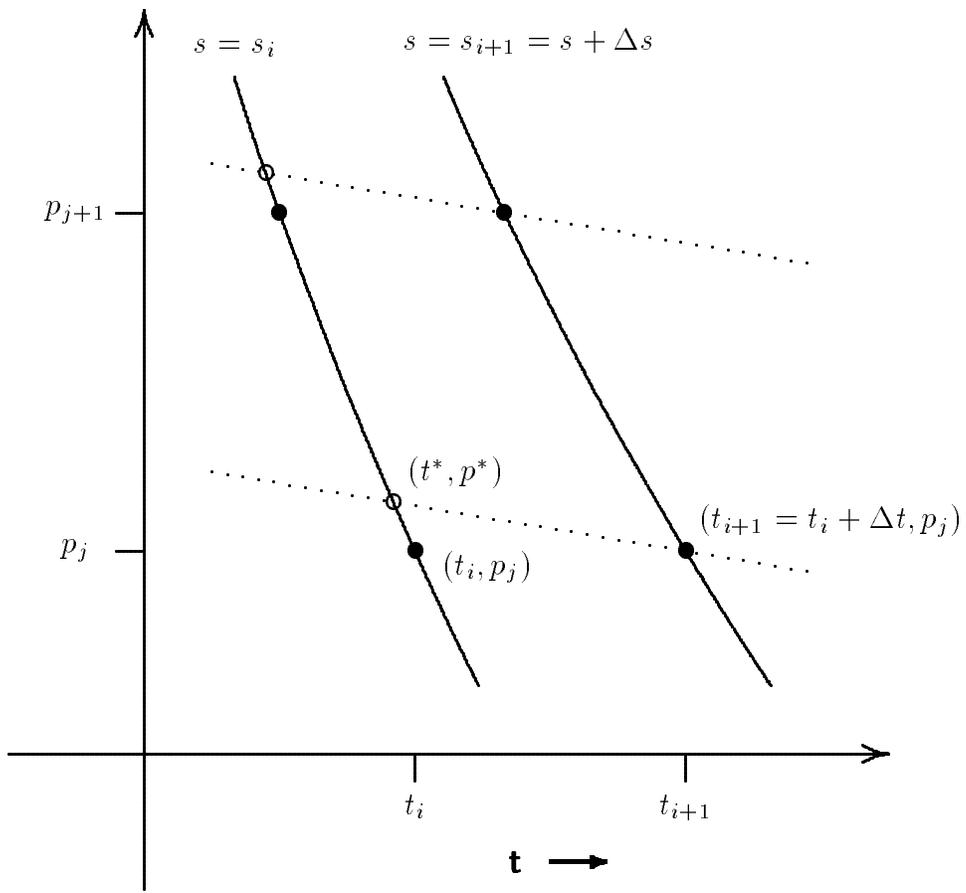
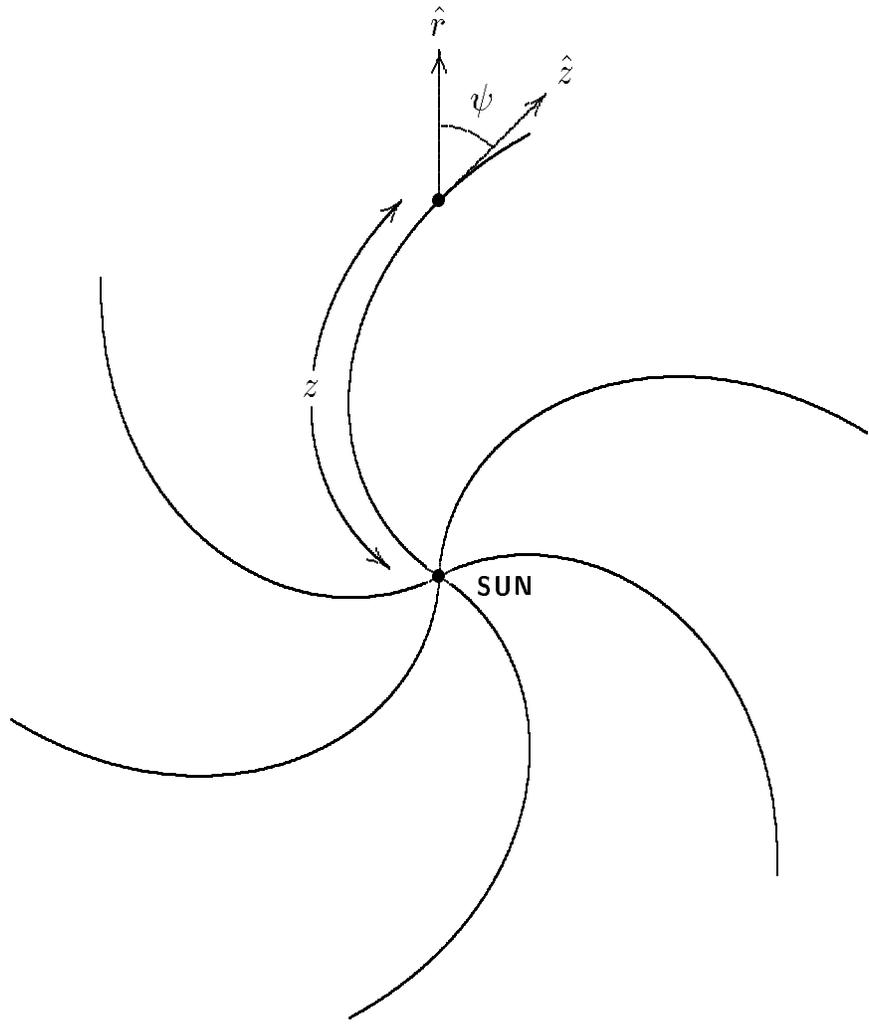
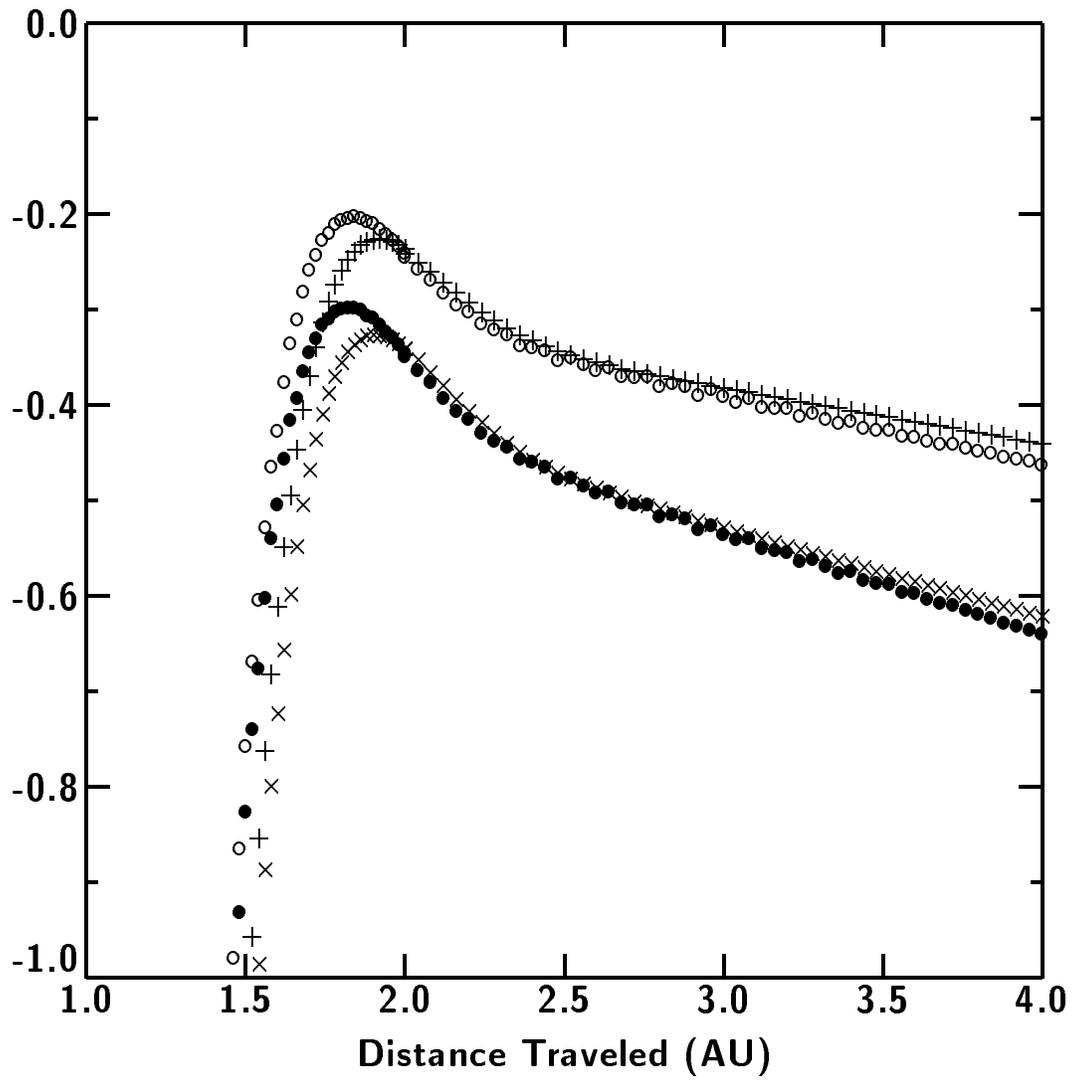
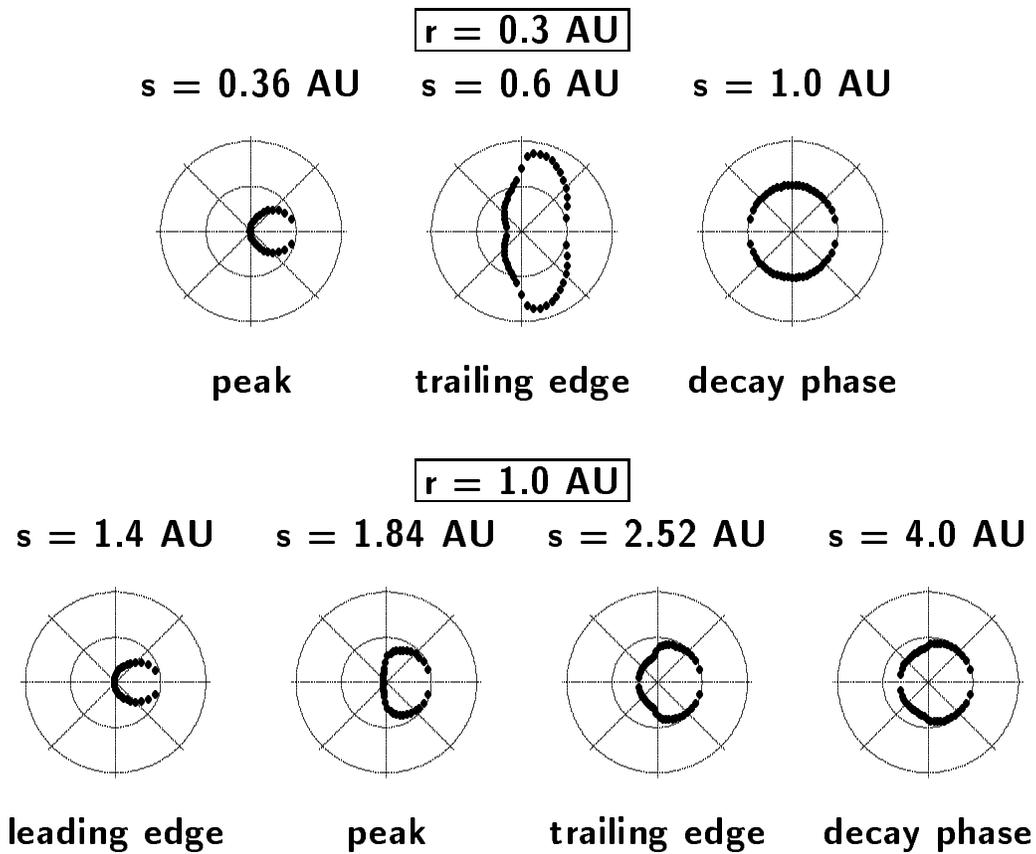
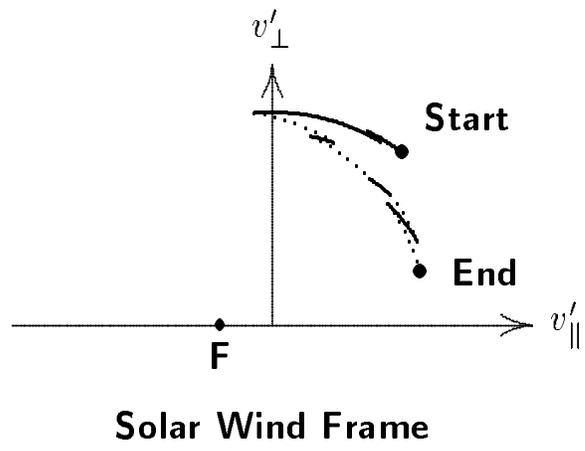
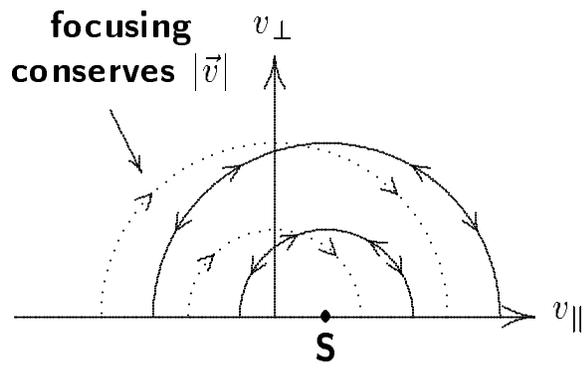
\bye